\documentclass[]{article}

\def\ben{\begin{equation}}
\def\een{\end{equation}}
\def\bea{\begin{eqnarray}}
\def\eea{\end{eqnarray}}
\input amssym.def
\input amssym.tex
\begin{document}

\hfuzz=100pt
\title{Thoughts on Tachyon Cosmology}
\author{G.W. Gibbons
\\
D.A.M.T.P.,
\\ Cambridge University,
\\ Wilberforce Road,
\\ Cambridge CB3 0WA,
 \\ U.K.}
\maketitle

\begin{abstract}
After a pedagogical review of elementary cosmology,
I go on to discuss some obstacles to obtaining
inflationary  or accelerating universes in M/String Theory.
In particular, I  give an account of an old ``No-Go Theorem''
to this effect.
I then   describe some recent ideas about the
possible r\^ole
of the tachyon in cosmology. I stress that
there are many objections  to a naive inflationary  model
based on the tachyon, but
there remains the possiblity that the tachyon
was important in a possible pre-inflationary
``Open-String Era '' preceeding our present ``Closed String Era''.

\end{abstract}

\vfill \eject

\section{Introduction}

The  original aim of my lecture was to give a brief outline some recent work
and its  justification on the possible
role of the tachyon in cosmology \cite{roller}. I also wanted to describe some
recent  work
with Hashimoto and Yi \cite{GHashimotoY},
 extending some older work with Hori and Yi \cite{GHoriY}
on the  effective-field-theoretic  details of the process of tachyon
condensation.  In particular I intended to describe what we have
called the Carollian confinement mechanism for open string states.
Since this entails familiarity with the Carroll group
I planned to include some some new and old material on that as well.

However I was also asked to cover some more elementary topics
in cosmology by way of an introduction. This I have
done, but it has lead to  a slightly lengthy write-up.
Since the subject of cosmology is  not
short of excellent  books and reviews,
I have approached the subject from a deliberately
idiosyncratic fashion which I hope may appeal
to those whose appetite  has
become somewhat jaded while not misleading the beginner.
The reference list, especially when it comes to elementary cosmology,
is rather selective, not through any desire
to slight the many valuable contributions to the subject
but to limit what would be a very large list
to  a size  which I could handle.

\section{Elementary Cosmology}

Before getting on with the main business, I will,
as requested by the organisers and various participants,
begin  with a summary of elementary
cosmology. As a first step, we assume the {\it Cosmological Principle}
which states that the universe is
spatially
homogeneous and  isotropic. Observational  support for this
assumption
comes both from the high degree of isotropy of the cosmic microwave
background
radiation (CMB), radio observations and the large scale  distribution
of galaxies.

\subsection{Kinematics of Homogeneity and Isotropy}
Mathematically we reformulate the assumption as stating that
 locally  the metric admits
a six-dimensional isometry group acting on three-dimensional
spacelike orbits $G/SO(3)$ , with $G=SO(4)$, $E(3)$ or $SO(3,1)$.
These possibilities are indexed by a quantity $k$ taking the values
1,0,-1, respectively. Locally, the orbits, i.e. the spatial cross sections,
inherit the geometry of $S^3$, ${\Bbb E} ^3$ or $H^3$ respectively
with their standard metrics $ds^2_k$.
The metric thus takes the Friedman-Lemaitre-Roberson-Walker (FLRW)
form
\ben
-dt^2 + a^2(t) ds_k^2 ,
\een
where the function $a(t)$ is called the {\it scale factor}.

Globally it  is always possible to make
identifications on the orbits under the action
of a discrete sub-group $\Gamma \subset G$, even to the extent
that the orbits,
$\Gamma G/SO(3)$,
in the case $k=0,1$ become compact.  Therefore one should
avoid labelling the cases $k=1$ ```closed'' and $k=0,-1$ ``open''.
Future observations using the MAP satellite
will tell us
the extent to which these possibilities are actually realised \cite{CQG}.
In general such identifications destroy  isotropy and
 also homogeneity. The latter may be avoided by taking $k=1$
and  identifying
$S^3$ with $SU(2)$ and taking $\Gamma$ to act exclusively
on the left or on the
right. One may retain both isotropy and homogeneity by
taking $\Gamma$ to be the antipodal map. The spatial sections are then
${\Bbb R}{\Bbb P} ^3 \equiv SO(3) \equiv SU(2)/{\Bbb Z}_2$.
This is sometimes called the {\it elliptic}  case and was first
suggested
in the context of relativistic cosmology by Felix-Klein.
In fact in 1900,
long before the advent of General Relativity, the observational
consequences of the universe being spatially $S^3$ or ${\Bbb R} {\Bbb
  P}^3$
had been investigated by Karl Schwarzschild (see \cite{CQG}
for an English  translation of his paper).

In a general FLRW spacetime  the Lorentz  group
is spontaneously broken down to it's orthogonal subgroup.
The cosmological background acts in this respect  as a kind of \ae ther.
This ``\ae ther'' is translation invariant, but if $k\ne 0$,
the translations do not commute.

 For particular scale factors however the
metric  admits {\it enhanced symmetry}:
there are  additional isometries
rendering it homogeneous both in space and
time and in fact locally  maximally
symmetric and thus of constant curvature
\ben
R_{\mu \nu \alpha \beta }= { \Lambda \over 3}  \bigl (g_{\mu \alpha }
g_{\nu \beta} - g_{\mu \beta} g_{\nu \alpha}   \bigr ).
\een
This  implies that the metric is also Einstein
\ben
R_{\mu \nu} = {\Lambda \over 3} g_{\mu \nu}.
\een
The cases are
\begin{itemize} \item de-Sitter spacetime $dS_4 \equiv
  SO(4,1)/SO(3,1)$
which happens if $k=1$ and $a(t) = \cosh(Ht)$, $k=0$ and
$a(t)= \exp (Ht) $, and $  k=-1$ and $a(t)=\sinh(Ht)$, with $
H^2 = {\Lambda \over 3} $
\item Minkowski spacetime ${\Bbb E} ^{3,1} = E(3,1)/SO(3,1)$, which
  happens if $k=0$ and $a(t)= {\rm constant}$ or $k=-1$ and $a(t)=t$.
\item  Anti-de-Sitter spacetime $AdS_4 \equiv SO(3,2)/SO(3,1)$,
which happens if $k=-1$ and $
a(t) = \sin (Ht) $ with $H^2 = - {\Lambda
  \over 3}$.
\end{itemize}

Some of coordinate systems are purely local.
For example the {\it Milne  universe} (i.e. the $  k=-1$ form
of Minkowski spacetime)  covers only  the interior of the future
light cone of a point. The light cone corresponds to the
coordinate  singularity  at $t=0$. This model was originally
introduced to describe a single creation event for the universe
avoiding the use of general relativity.
The  {\it Steady State universe} was invented to achieve the opposite
goal. However it
(i.e. the $k=0$ form of de-Sitter spacetime) is geodesically
incomplete in the past and covers only
half of de-Sitter spacetime. There is a
past horizon at the coordinate singularity at $t=-\infty$.
Thus it is really not past eternal as the founders of that theory had
supposed and indeed there are now grounds for doubting
that inflation can ever be past eternal \cite{bgv}.

In connection with eternal inflation, one might
ask what 4-dimensional Lorentzian
Einstein spaces can be homogeneous, that is admit
a transitively acting isometry group. This requirement is a weak
version
of what Bondi calls the {\it Perfect Cosmological Principle}.
The list is not large
and moreover contains a surprise. In addition to
the previously noted maximally  symmetric cases,
Minkowski spacetime, ${\Bbb E}^{3,1}$, De-Sitter spacetime $dS_4$ and
Anti-de-Sitter spacetime $AdS_4$, there are the two expected two products
$dS_2 \times S^2$ and $AdS_2 \times H^2$. If the cosmological constant
is positive, $\Lambda >0$ this is all
there are. If $\Lambda \le 0$ there are two more cases. If $\Lambda=0$
there are homogeneous plane waves, some of which are geodesically
complete
and another case of Petrov type I.
If $\Lambda <0$ one has a Petrov Type N generalized plane wave
and another Petrov type III solution.

Thus if eternity  is homogeneous and the cosmological constant is
positive,
then there is really only de-Sitter spacetime or $AdS_2 \times S^2$
to choose from.

Just as the Poincar\'e group contracts to the Galilei group
or the Carroll group \cite{Leblond, Sen},
the de-Sitter and Anti-de-Sitter groups have interesting
contractions which have been classified in \cite{Bacry, Nuyts}.
One may also obtain the Poincar\'e group as a contraction
of the de-Sitter or Anti-de-Sitter groups and these are in fact
the only groups for which this is possible \cite{Levy-Nahas}.

\subsection{Kinematics of the Energy Momentum Tensor}

It follows directly from the metric and the Einstein equations
or, if one wishes, just  from the symmetry assumptions,
that the energy momentum tensor
takes the form of a {\it perfect fluid  }
\ben
T_{\mu \nu} dx ^\mu dx ^\nu = \rho(t) dt ^2 + P(t) g_{ij} dx ^i dx ^j,
\een
where $\rho(t)$ is called the energy density and $P(t)$ is called the
pressure. The {\it Weak Energy Condition}  requires that the energy
density is non-negative
\ben
\rho \ge 0.
\een

Eliminating the coordinate $t$ leads to an {\it equation of state}
$P=P(\rho)$. Note that this is purely a kinematic statement
without necessarily having any  thermodynamic content.
{\it A priori}, the pressure need not even be a single valued function
of density. If
\ben
P= (\gamma -1) \rho,
\een
with $\gamma$ a constant one refers to a {\it polytropic  equation of
 state}. More generally one may define the ratio
\ben
w= {P \over \rho},
\een
and in the polytropic case $w=(\gamma-1) ={\rm constant}$.

The speed of acoustic or sound waves
 $c_s$ is given by Newton's formula
\ben
c_s^2 = {\partial P \over \partial \rho}.
\een
It is real and less than that of light if
\ben
1 \le   \gamma \le 2.
\een
If $\gamma =1 $, i.e.  $P=0$,  one speaks of pressure free matter or `dust'',
 $\gamma = { 4 \over 3}$, $  P= { 1 \over  3} \rho $,
 one speaks of ``radiation" and if $ \gamma
 =2$, i.e. $P= \rho$  one speaks of ``stiff matter''. Thus  acoustic
waves in a radiation fluid   travel at a
speed $ 1 \over \sqrt 3$. This  is  the analogue for a gas of photons
of ``second sound"  in a gas of phonons. The acoustic peaks in the
BOOMERANG observations of the CMB
show clear  evidence for the propagation of  ``second light''.

The {\it Dominant Energy Condition}
 allows a negative pressure as well as
the usual positive pressure  as long as
\ben
|P| \le \rho.
\een
As an example consider a single scalar field $\phi$  with potential
$V(\phi)$. If $\phi$ depends only on time then
\ben
\rho = { 1\over 2} {\dot \phi} ^2 + V(\phi),
\een
\ben
P= {1\over 2} {\dot \phi } ^2 - V(\phi).
\een
If the  kinetic energy dominates we have stiff matter, $P=\rho$,
with the maximal possible positive pressure  .
If the potential energy dominates  and we have the greatest  possible
tension  or negative pressure $P=- \rho$.

\subsection{Consequences of the Einstein equations}

The Einstein equations provide three equations.
\begin{itemize}
\item{\it Raychaudhuri's equation}
\ben
{ \ddot a \over a} = - { 4 \pi G \over 3} ( \rho +3 P) \label{raychaudhuri}.
\een

\item {\it Friedman's equation}
\ben
{\dot a \over a} ^2 + { k \over a^2 } = { 8 \pi G \over 3} \rho.
\label{Friedman} \een

\item{\it  The First Law}
\ben
{\dot \rho } + { 3 \dot a \over a} ( \rho +P)=0 \label{First}.
\een

\end{itemize}

The Bianchi identity implies that of these only two are independent.
The ``First  Law``  expresses the fact that the  energy
in a co-moving volume  $a^3 \rho$
is not conserved
during the expansion of the universe. If one assumes that the matter
is in local thermodynamic equilibrium and passes adiabatically
through a succession of
equilibrium states whose  equilibrium pressure
is given exactly by $P$, then one may deduce that the entropy
in a co-moving volume  is conserved. In practice the matter is never
in complete equilibrium, moreover it is always tending to the
equilibrium configuration of constant entropy  at a  lower and lower
energy, i.e at a  lower and  lower temperature.
This is why naive
nineteenth century ideas about the `` Heat a Death of the Universe''
were essentially misguided. They were predicated on the assumption
that
the matter in the universe  was both  static and  isolated.

If the  equation of state  is known one may integrate  the first law to
obtain
the pressure as a function of the scale factor. In the polytropic case
one obtains
\ben
P \propto { 1 \over a ^{ 3 \gamma } }.
\een
From this it follows that if the universe is expanding then the
density of radiation falls of with one more power of the scale factor
$a$
than pressure free matter, which will therefore ultimately dominate.

\subsection{Future and Past Horizons}
We now turn to light propagation and horizons.
If $r$ is the radial coordinate on  $S^3$,$ {\Bbb E} ^3$ or $ H^3$
with us taken as the origin, then
by symmetry, radial null geodesics   are given by $r=\eta$
with
\ben
d \eta = { dt \over a(t)}  ,
\een  The coordinate $\eta$ is called {\it conformal time} because
using it, the FLRW metric becomes explicitly conformally static
\ben
a^2 (t) \Bigl  ( -d\eta ^2 + g_{ij} dx ^i dx ^j \Bigr  ).
\een
In fact the FLRW metric is also conformally  flat
but we shall not need that fact here.

 To read of the causal structure
it suffices to consider the  {\it Penrose diagram} i.e.
the orbit space of the  $SO(3)$ action,
 which has coordinates $r \ge
0$
and $\eta$ in which light rays are at 45 degrees.
Thus if the polytropic index is a constant so that
$\rho \propto{ 1  \over  a^{3 \gamma}}$ we define $y=a^{3 \gamma -2 \over 2   }$
\ben
{ d^2 y \over d \eta ^2 } +k({3 \gamma \over 2}-1)^2 y=0.
\een
If $k=1$, this is a simple harmonic oscillator with total duration
from Big Bang to Big Crunch equal to $2\pi \over |3 \gamma -2  |$.
Thus for dust, $\gamma=1$, a lightray will just circumnavigate
an $S^3$ in that time, while for de-Sitter spactime with $\gamma =0$,
it would only get half way. The same is true of  radiation
for which $\gamma ={4 \over 3}$, while for stiff matter, $\gamma =2$,
it only gets one quarter of the way around.

The structure of the conformal boundary is determined
by the range of $\eta$.  As originally clarified by Rindler,
\begin{itemize} \item If the universe expands  forever and  :
\ben
r_+( t) = a(t) \int ^\infty _t { dt^\prime  \over a(t^\prime )} < \infty,
\een
there will be a {\it future event horizon}, that is a null hypersuface
separating events
which the observer at $r=0$ will see from those he or she  will
never see. In this case the future boundary at $t=\infty$ is spacelike
.The standard example of a cosmological spacetime with
a future horizon is de-Sitter spacetime. In this case the
(observer dependent) horizon
is also  Killing horizon  and its area $A_+(t)= 4 \pi a^2(t) r_ + (t)$
is constant with time. In general, if the weak energy condition
 holds,
$A_+(t)$ is non-decreasing. In the static  case
one  has an associated constant temperature $T_+= { 2 \pi \over H}$.
It is reasonable to assume that if $A_+(t)$
does not change too fast one simply replaces $ 1\over H$
in this expression by the instantaneous Hubble radius.
One usually ascribes a non-decreasing  entropy
\ben
S(t) = { 1\over 4 G} A_-(t),
\een
to each co-moving  observer's horizon and interprets it
as  a measure of the total  information lost to he or she
up to time $t$.

\item If the scale factor vanishes at $t=0$, a  {\it Big-Bang }
singularity, and
\ben
r_ -(t) =a(t)  \int _0 ^t  { { dt^\prime}   \over { a(t^\prime )} } <
\infty ,
\een
then there will be a {\it  past  horizon }, that is a null hypersurface
separating events with which an
observer at $r=0$ can have been in causal contact since the Big Bang
from those he or she
cannot have been. In this case there is a singular spacelike past
boundary at $t=0$.  Events with time coordinate $t$ and radius $r< r_-(t) $
are said to be inside  the {\it particle horizon}. If identified with
the points on co-moving geodesics for which $r$ is independent of time,
the entire past  geodesic is outside the past light cone of the
observer at the origin. Indeed a light ray  starting from $t=0$
and $r=0$  just reaches  the particle horizon at any subsequent time.

\end{itemize}
Note that to calculate the particle horizon radius $r_-(t)$ at time
$t$ requires knowledge of the scale factor at all previous times.
In other words it is an anti-teleological concept, just as the future
horizon is a teleological  concept. It is a widespread, misleading
and incorrect
practice
to muddle the particle horizon  with the more or less directly
  observable
and instantaneously defined
{\it Hubble radius}
\ben
r_H(t) = {a(t) \over {\dot a} (t) }.
\een
If the scale factor goes to zero  like $t^p$ near $t=0$,
then there will be a particle horizon if $p <1$. For a $k=0$
universe with a polytropic equation of state, the Einstein equations
imply
\ben
p= { 2 \over 3}.
\een
Regardless of the actual value of $k$,  it is almost always a good
approximation  to set it to zero near  a  Big Bang because the term
$k \over a^2$ in the Friedman equation (\ref{Friedman}) is negligible
compared with the other terms.

\subsection{ The Horizon Problem}
This gives rise to the {\it Horizon Problem}. The point is that
the present  day observed CMB photons at temperature roughly
 3K  arrive unscattered from the spherical  intersection $S_{ls}$  of
our past light cone with the
{\it Hypersurface of  Last Scattering}   $t=t_{ls}$, red shifting by factor
$1+z= {a(o) \over a(t_{ls} } ) \approx 10 ^3$ on the way. Note that they
  keep
their almost perfectly  thermal spectrum purely by means of the
kinematic
effect of their redshift. Before last scattering $t=t_{ls}$ they
are presumed to have been in thermal contact. Observations of the CMB
reveal that the temperature on $S_{ls}$  was uniform
to better than one part in $10^4$. This uniformity is a puzzle if one
assumes
that the universe was radiation dominated all the way back to a Big
 Bang before $t=t_{ls}$  ( i.e. $ p= { 1 \over 2})$   because
the horizon radius $r_-(t_{ls}) $  calculated under that assumption is
about 80 times smaller than the proper diameter $2 a(t_{ls})  r_{ls}$ of
the last scattering sphere $S_{ls}$. In other words, under this
assumption,  the past light cones of most  pairs of points   on
  $S_{ls}$ never intersected.

The recent BOOMERANG and MAXIMA data on the deviations of the temperature
of the CMB from isotropy allow us to sharpen this argument.
Roughly speaking, such second light behaves like
\ben
{ \delta T  \over T} = \int d^3 {\bf k} \thinspace
 e^{i{\bf k} .{\bf x}} \bigl [
A({\bf k}) \cos ({ k \eta \over \sqrt 3}) + B({\bf k}) \sin ({k \eta \over
  \sqrt 3}) \bigr ].
\een
The cosine mode is known to cosmologists  as the ``iso-curvature"
 mode and the sine mode as the ``adiabatic'' mode.
The physical distinction between them is that in the latter case
the ratio of baryon number to photon number is independent
of position whereas in the former it depends upon position.
Moreover for ``causal''  perturbations one expects that
the cosine mode is absent because it varies non-trivially at the
largest scales, i.e. at the smallest possible values of $k=|{\bf k}|$.
If one decomposes the temperature $T=T(\theta, \phi)$
on the sphere of last scattering $S_{ls}$
into spherical harmonics
\ben
{ \delta T \over T}  = \sum a_{lm} Y^m_l(\theta, \phi),
\een
one finds so-called  peaks in the amplitudes, the lowest  at $l \approx 200$.
These can only be explained if the sine mode is absent, thus
establishing
that the primordial fluctuations have scales much larger than
could have arisen from causal processes if the past horizon
was roughly the Hubble radius at that time.

A more detailed examination of the data reveals more.
The universe is almost flat and the content roughly
$30\%$  cold dark non-baryonic matter, and $70^\%$
 matter with negative   pressure
and $w< -.5$.

\subsection{Inflation}
The theory of {\it Inflation}
seeks to remedy this, and other defects of the {\it Standard
  Cosmological Model}  by postulating a much more rapid growth of the
scale
factor sometime before $t_{ls}$; growth at least  as fast as $t$.
If true, this means  that the scale factor must have accelerated
some time before $t_{ls} $,
\ben
{\ddot a} >0.
\een   It follows from the Raychaudhuri equation,
that the pressure must have been at least as negative
as $-{ 1\over 3 } P$. In other words,  the universe must have
{\sl anti-gravitated} in the past. Using
\ben
\rho + 3 P = 2 \bigl ( T_{00} - { 1\over 2} g_{00}
 T^\lambda \thinspace _\lambda \bigr )
= { 1\over 4 \pi G} R_{00},\een
the Raychaudhuri equation
(\ref{raychaudhuri}) may be written as
\ben
{{\ddot a} \over a} = { 1\over 3} R_{00},
\een
we need
\ben
R_{00} <0.
\een
In the language of energy conditions, we need a massive
violation of the {\it Strong Energy Condition} in the past.

Observations of distant Type 1A supernovae, which
essentially use the curvature of the redshift magnitude or {\it Hubble
diagram},
assuming that this class of  super-novae provide us with
standard candles whose light is unscattered on its way to us,
also seem to indicate that
the universe is accelerating at the present epoch. To explain this
anti-gravitating behaviour one is  forced to accept directly from
Raychaudhuri's equation (\ref{raychaudhuri}), which is no more than a
statement of Newton's law of gravitation,
that there must be
massive violation of the strong energy condition, with large negative
pressures
at late times. Indeed if one models  the data one finds
that about 70\% of the energy density is of this exotic form,
the bulk of the remaining matter is
dark, non-baryonic  and   pressure free and
less than 4 \% is ordinary baryonic matter.

 \section{The Inflaton}

The standard way of arranging for
a violation of the strong energy condition
is to  postulate the existence of one or more hitherto unobserved
scalar fields admitting everywhere {\sl non-negative  potentials}
usually with an absolute  minimum for which the potential
either vanishes or is very small. Because potentials have dimensions
mass to the power 4, what is required to get a suitable non-vanishing
value at late times  is a mass scale near the minimum
less than $10
^{-4} eV$. Curiously, this is rougly the order of magnitude of
some neutrino masses, in  fact which perhaps lends greater credence
to the idea.

The simplest case is just one scalar field, called in this context
 an {\it inflaton}. It is usually assumed to start near the top of a
 fairly
flat potential and to roll slowly down to the minimum of the
 potential.
The potential needs to be fairly flat so that sufficiently many
 e-folds ( at least 60) take place before the minimum is reached.

During slow rolling the  second derivative term in the equation of motion
\ben
{\ddot \phi} + { 3 {\dot a} \over a} {\dot \phi} + {\partial V \over
 \partial \phi} =0,\label {slow}
\een
is neglected so
\ben
 { 3 {\dot a} \over a} {\dot \phi} + {\partial V \over
 \partial \phi} \approx 0,\label {slower}
\een
and  the Friedman equation is
approximated by
\ben
{{\dot a}^2  \over a ^2 } \approx { 8 \pi G \over 3} V(\phi) \label{slowest}.
\een
Given $V(\phi)$ one may now solve
(\ref{slower}) and (\ref{slowest}) in what is called {\it the Slow
 Rolling  Approximation}
to find $a(t)$. I will not give the easily derived
formulae here. It is important to
realise that because $V(\phi)$ is not constant during  slow rolling,
the inflation is not exponential and so the de-Sitter metric
with $H=\sqrt{ 8 \pi G V(\phi) \over 3}$ is only an  approximation
to the actual metric at that time.

An interesting  simple case for which an exact solution is available
is given by \cite{Linde}.
\ben
V=V_0 \exp (\alpha \phi) .
\een
If $k=0$, one has a  solution with
\ben
 V \propto { 1 \over t^2}, \qquad a(t) \propto t.
\een
In string theory, one may think of $\phi$ as the dilaton.
The solution is then a timelike version of the usual
spacelike linear dilaton vacuum. In string conformal frame,
the metric is flat.

\subsection{``Reheating ''}

Slow roll over is usually supposed to terminate with a
period during which the inflaton executes damped  oscillations
near the bottom of the potential. During this period its coupling to
the matter of the {\it Standard Model of Particle Physics}
leads to something called in the characteristically
confused and muddle
and misleading jargon of cosmologists  which we encountered
when applied to the Hubble radius, ``re-heating ''
since it is not clear, nor more damningly very relevant,
whether the universe was
ever  heated before this time. (Similar
strictures
apply to the terms ``recombination'' and ``adiabatic '').

Polemic aside, the main point is to get some of the energy
in the  oscillations of the inflaton to produce
a radiation dominated universe at a suitably high temperature
so that baryon-violating processes can give rise to the puzzling
small number  of
baryons that we both see and love today
compared with the very  large number of photons
in the CMB.

\subsection{Planck Scale Inflation}

To get a suitable inflationary scenario, not only must one
assume that the potential $V(\phi)$ is such as
to ensure sufficient inflation, one also needs to ensure that
the transition to a radiation era is not too abrupt,
otherwise more gravitational waves  will be produced
than are compatible with the isotropy of the CMB.
This is usually taken to bound below  the
value of the Hubble radius during inflation. If $m_p$ is the Planck
mass
then one needs roughly
\ben
{1 \over H } > 10 ^{-5} {1  \over m_p} .
\een
In other words naive models involving Planck scale inflation are
very probably ruled out \cite{Stephen}.

\section{Machian Considerations, Chronology Protection
and the Spontaneous Breakdown of Lorentz-Invariance}

Many cosmologists in the past have worried about
``Mach's Principle''.
This can be taken in many ways. I prefer
to think of the `` Machian Coincidence ''.
 This is the fact that locally,
without reference to any astronomical
observations,  one may define
a non-rotating inertial frame, i.e  one in which the usual laws of Newtonian
mechanics hold,  while astronomically we may define a frame
with respect to which distant glaxies are at rest and one finds
that these two frames coincide. Traditionally the first frame is
determined using a version of Newton's bucket experiment.
The second
may nowadays be determined in terms of the last scattering surface
rather than the distant stars or galaxies.

The basic example of a spactime in which these two
standards of rest disagree is  the
well known  Goedel solution.  It  is homogeneous,
and hence stationary. This  universe, which is spatially non-compact,
 rotates about every point. However
it admits closed timelike curves. A less well-known but in some ways
much
more striking example was constructed by Ozsvath and Schucking
\cite{mach}.
It is also homogeneous but the space sections are squashed $S^3$'s.
More importantly, it is causally well behaved. In particular
it has no CTC's.

Some years ago, Hawking \cite{Hawkingrot},
 by considering the transverse Doppler effect,
argued that the extent to which the two frames agree
 is such  that the universe
can have completed only a very small, fraction of a rotatation
since the epoch of combination at
which the cosmic plasma became transparent. The original bounds of
Hawking have been strengthened somewhat  by Barrow, Juskiewicz and
Sonada \cite{JBS}.

One mathematical formulation of the Machian coincidence
is to consider the average  4-velocity $\bf U$ of the matter
in the universe. This may be defined as the timelike
eigen-direction of the energy momentum tensor. Inertial frames
are Fermi-Walker propagated along the integral curves of $\bf U$.
Given the vector field $\bf U$ one may also define
its vorticity $\omega
= \star u \wedge du$, where $u$ is the one form obtained from
the vector field $\bf U$ by lowering an index using the metric.
The Machian Coincidence is essentially the fact
that our present vorticity is very small.
One   explantion for this fact comes from Inflation
as pointed out by Ellis and Olive \cite{EllisOlive}.
Their point is that  as the universe expands, then by angular
momentum conservation, the vorticity necessarily decreases.
Moreover according to inflation,
 there was probably no initial vorticity
to begin with. The first  argument holds in almost all
inflationary scenarios and the second also holds in inflationary scenarios
based on single scalar field.
In that case vorticity perturbations necessarilly vanish.

As an aside, let me remark that
a striking feature of of many spacetimes with non-vanishing vorticity,
which has been known almost since the beginnings of general
relativity,
 is the existence in them of closed timelike curves.
Because of the expansion of the universe, violations of Mach's
coincidence now require large vorticities at early times.
If those large vorticities entailed the existence of CTC's
and some mechanism, along the lines of that invoked
in   Hawking's Chronology Protection
Conjecture holds, that might have
prevented the occurrence of CTC's thus providing  an alternative
explanation: the large vorticiy could not have been generated in the
first place. For further references and
a   calculation supporting this idea see \cite{Huang}.

The absence of vorticity implies,  a velocity
potential, and hence, if the fundamental group of spacetime is trivial,
 a time function, i.e. a function which increases
along every timelike curve. The existence of such a
function, which is defined
only up to a re-parameterization, guarantees that there are no CTC's.
The time-function also provides an absolute rest frame or \ae ther.
It may be thought of as the analogue of the Higgs field, responsible
for the spontaneous breakdown of Lorentz-invariance, itself somewhat
of a Machian conundrum.
 A time function
is  also what is needed
to resolve  the much discussed  {\it Problem of Time} in quantum cosmology.
In inflationary scenarios these functions are performed by
the mysterious inflaton. In the theories we are about to describe
it is the tachyon field which is responsible. For some recent
ramifications of this idea, the reader is directed to \cite{Sen2}.

\section {M-Theory and Cosmology}

The elementary introduction makes it clear that
if we are to make contact between cosmology and
fundamental theory, such  as M/String Theory
we need to violate the strong energy condition and
more particularly, obtain scalar fields with positive potentials.
This is a notoriously difficult thing  to do.  It is not just that
de-Sitter spacetime is not a supersymmetric background
and so reliable quantum calculations  like the those around
Anti-de-Sitter spacetime cannot be done.
The problem is much more basic than that. It is
that  pure supergravity theories, without additional
supermatter in arbitrary dimensions  satisfy the strong energy
condition
and thus do not permit accelerating behaviour.
As a consequence they also do not allow oscillatory
or cyclic behaviour for the scale factor. In fact for compact
universes, Tipler has adapted  the techniques used
to prove the Hawking-Penrose Singularity Theorems
to show that  the strong energy condition rules out
any type of recurrence
behaviour \cite{Tipler}
. For related reasons, the negative tension branes required
in certain brane-world scenarios violate  even the weak positive energy
condition and cannot come out
of supergravity theories \cite{GibbonsKalloshLinde}.

As is well known, if a cosmological term is possible
it must be negative, not positive. The reasons for this seem to be
rather deep. One explanation comes from thinking of the theories
in higher dimensions, specifically 10 and 12 dimensions.
This will be the subject of the next section.
It is well known that gauged supergravities in 4 spacetime dimension
for vector  gauging  have negative cosmological constant
\ben
\Lambda =  - {3  e ^2 \over 4 \pi G}.
\een
Curiously for axial gauging Freedman \cite{Freedman} showed that
one has a positive cosmological constant
 \ben
\Lambda =  - (\gamma _5) ^2 {3  e ^2 \over 4 \pi G}.
\een
However such axial gauging leads to anomalies
and hence to non-viable theories. It seems that
this problem can arise in hybrid inflation models  in which
Fayet-Illiopoulos terms come into play.

\section{Compactification and a No-Go Theorem}

In what follows, I shall recall some facts
which have been known since the early '80's but which still seem
relevant today \cite{Gibbons4}. Despite the passage of  time,
compactification
is still a very imperfectly understood  process from the physical
point of view.
One speaks of an $N$-dimensional     spacetime
$M$  being compactified
if it is a (possibly  warped) product of    a Lorentzian
spacetime $X$ with a Riemannian spacetime $Y$:

\begin{eqnarray}
M=X \times_W  K. \label{compact}
\end{eqnarray}
with a metric given in local coordinates $x^M$ which split as
$x^\mu$  for $X$ and
$y^m$ for $Y$
\begin{eqnarray}
g_{MN} dx^M dx ^N =
W^2 ( y) g_{\mu \nu }(x) dx ^\mu dx ^\nu + g_{mn} (y) dy^mdy ^n,
 \label{metric}
\end{eqnarray}
where $W(y)$ is called the warp factor. This statement  gives no insight
into the process of compactification itself, which was presumably
a dynamical process which occurred in the early universe, possibly as
some sort tunnelling process ``{\it ex nihilo}  or as a semi-classical
approximation to the wave function of the universe according to the ``no
boundary" proposal. An alternative picture is obtained if one
views the present universe as
the outcome of some sort of
dynamical collapse to a branelike  configuration. More generally one
might envisage that $M$ contains  more than one brane which may also
undergo collision. In the former two
cases at least
it is natural to assume that $Y$ is compact, complete and non-singular
 without boundary $\partial Y= \empty 0$ and that the warp factor
$W(y)$ is a smooth and no-where vanishing function on $Y$.
In the latter cases, $Y$ may well be non-compact and $W$ might well
vanish at the location of Killing horizons.

Rather  general restrictions on the form that the metric
 (\ref{metric})
may take are obtained if one recalls (or checks that) that for all pure
supergravity models the bosonic energy momentum tensor $T_{MN}$
satisfies
the Strong Energy Condition.
The Einstein equations
in $M$ read:
\begin{eqnarray}
R_{MN}= {8 \pi G_N} \bigl ( T_{MN} - { 1 \over N-2} g_{MN} { T ^L}
\thinspace  _L
\bigr) \label{einstein}.
\end{eqnarray}
and the strong energy condition is the statement that
for all non-spacelike vectors $T^M$

\begin{eqnarray}
R_{MN}T^MT^N\ge 0,
\end{eqnarray}
 or in any local coordinate system
\begin{eqnarray}
R_{00}\ge 0.
\end{eqnarray}
Physically the Strong Energy Condition is the condition that locally
gravity is attractive. It amounts to saying that any tensions or
negative pressures, themselves a rather unusual occurrence,
can never exceed ${ 1 \over N-1}$ times the energy density.
In other words, a medium not satisfying the Strong Energy Condition
must be  one supporting extremely large tensions or negative
pressures. Usually such a medium  would be expected to be
highly unstable. It is curious that in pre-relativistic
discussions of Newtonian Gravity, a stress tensor was ascribed
to the \ae ther with precisely
this odd property (see \cite{Max}) and it was felt, for example by
Clerk Maxwell, to be odd at the time \cite{Synge}.

 It is not immediately obvious why
this restriction (considered via \ref{einstein}) as a condition on
the energy momentum tensor is satisfied for {\sl all} by  all the usual
bosonic energy momentum  tensors
considered in physics, built  for example from
p-form field strengths, $p \ge 2$, and for minimally coupled
 scalar fields
with the exception of
scalar fields with a
potential which can somewhere become positive. Moreover the
inequality is strict in the sense that
if $R_{00}=0$ then the energy momentum
tensor must vanish and hence the bosonic fields must actually  vanish.
For minimally coupled massless scalars without potentials, one finds
that they can depend on space but not on  time. In checking
the Strong Energy Condition in string theory,  it is essential to
work in Einstein conformal frame, in which the dilaton
is minimally  coupled. In string conformal frame, second derivatives of
the dilaton appear on the right hand side of Einstein's equations
and nothing can then be said about the sign of $R_{00}$.

It is important to note that
if the Strong Energy Condition is satisfied for the energy
momentum  tensors then it is also  satisfied for arbitrary positive
linear combinations. Thus for a multi- component system of fields,
possibly a statistical mixture, it is necessarily satisfied
if the individual  components of the mixture satisfy it.
For this reason it is satisfied for all the gaseous and liquid media
(so-called ``perfect fluids '')
that one typically considers in cosmology. It also implies
that the exponential factors involving  dilatons,
which appear ubiquitously in supergravity lagrangians,
do not prevent the Strong Energy Condition holding
if it holds for the individual pieces in the Lagrangian.

One should also  note that the Strong Energy Condition \cite{Hawking}
is independent of the Dominant Energy Condition. This states that in all
orthonormal frames $T_{00} \ge |T_{MN}|$, or more geometrically
if $T^M$ lies inside or on the future light cone then
so does $T^M \thinspace_ N T^N$. The Dominant Energy Condition
may be interpreted as requiring that matter may not move
superluminally.
It again is satisfied by all the usual bosonic energy momentum
tensors,
including minimally coupled scalars with {\sl positive } potentials.
Evidently it may  violated by minimally coupled scalars
with negative potentials. The Dominant Energy condition
is used to prove the Positive Energy Theorem  which in turn
implies that the purely  gravitational force between isolated systems
is always attractive.  In other words it rules out
the possibility of long range gravitational repulsions.
The Strong Energy Condition rules out antigravity
in a more local sense in that, as we shall see shortly,
it says that the source of the Newtonian potential
is  always locally of the same sign. Applied to cosmology it
implies that the acceleration of the universe is always  negative.
In particular it is incompatible with de-Sitter or de-Sitter-like
behaviour
for the spacetime metric $g_{\mu \nu} $ on $X$.

To see this in detail we use the formulae relating the Ricci tensor
$R_{MN}$
of $M$ to those, $^XR_{\mu \nu}$ and $^YR_{mn} $   of $X$ and $Y$
respectively. They are
\begin{eqnarray}
R_{\mu \nu}= { ^X R} _{\mu \nu} + g_{\mu \nu} { 1 \over p+1}
 { 1 \over W   ^{p+1}} \thinspace
 { ^Y \nabla ^2 }  ( W^{p+1} ) \label{Poisson},
\end{eqnarray}
and
\begin{eqnarray}
R_{mn}= { ^YR } _{mn} - { 1 \over W} \thinspace  ^Y \nabla _m \thinspace ^Y
 \nabla _n (W^{ p+1} ),
\end{eqnarray}
where $^Y \nabla_m  $ is the Levi-Civita covariant derivative on $Y$
and $^Y\nabla ^2 $ the Laplacian on $Y$ and the dimension of $X$ is
$ p+1$.
If $p=0$, then  we may with no loss of generality put $g_{00} =-1$, the warp
factor
is essentially the Newtonian potential of the static
$N$ dimensional metric on $M$ and (\ref{Poisson}) becomes the
relativistic
Poisson equation. If $R_{00} \ge 0$ then it will have a source with
the standard sign for a gravitational field. As a consequence the
total gravitational
mass of such a static spacetime  spacetime is non-negative.

Now let's allow $p >0$, and work in an arbitrary  orthonormal frame. We get
\begin{eqnarray}
R_{00}={ ^X R} _{00} - { 1 \over p+1} { 1\over W^{p-1}} ^Y\nabla ^2 (W^{p+1}).
\label{killer} \end{eqnarray}
If the ``internal `` manifold
$Y$ is compact and without boundary and $W$ smooth and nowhere
vanishing on it, then we may multiply equation (\ref{killer}) by $
W^{p+1}$
and integrate over $Y$. The term involving $^Y\nabla ^2$ vanishes
and we we deduce that

\begin{eqnarray}
^XR_{00}  \ge 0.
\end{eqnarray}
In other words, the Strong Energy Condition is hereditary. If it
holds on $M$ and we compactify on $Y$ then it continues to hold
on $X$. This is of course obvious for a strict metric
product for which the warp factor $W$ is constant. What we have shown
is that it remains true for a non-constant warp factor,
as long as $Y$ is
compact non-singular and without boundary.

If for example $X$ is a maximally  symmetric
spacetime  it must be an Einstein
space with $R_{\mu \nu} = \Lambda g_{\mu \nu}$ and we deduce
that necessarily the cosmological  constant is non-positive
\begin{eqnarray}
\Lambda \le 0.
\end{eqnarray}
Moreover if we want $X$ to be Minkowski spacetime  then
we can easily deduce from (\ref{killer}),
that since
\begin{eqnarray}
\int _Y W^{p+1}R_{00}  \ge 0,
\end{eqnarray}
 in fact $R_{00}=0$ and hence $R_{MN}=0$ and thus the higher
dimensional
spacetime must be Ricci flat. The bosonic fields must vanish.
It is now clear that $W^{p+1} $ is a harmonic function
on the closed manifold  $Y$ and hence it must be constant.
The compactification  must in fact be the metric product of Minkowski
spacetime with  a closed Ricci flat  manifold. Moreover
it follows  that, unless it splits as the (possibly local)
metric product of a circle
and
a lower dimensional Ricci flat manifold $Y^\prime$ , that the internal
manifold $Y$ can admit no continuous isometries. To see why this is
true,
recall that from Killing's equations, it follows that
 any Killing vector field  $K^m$  satisfies
\begin{eqnarray}
-^Y \nabla ^2 K^m - R^m \thinspace _n K^n =0 .\label{killing}
\end{eqnarray}
 Multiplying (\ref{killing}) by $K_m$ and integrating over $Y$
shows that $K^m$ must be covariantly constant,
\begin{eqnarray}
^Y \nabla _m K_n=0.
\end{eqnarray}
In particular we find that $g_{mn} K^m K^n $ is constant
and that $K_m =  {\partial \over \partial y^m}  \theta $ for some
function
$\theta$. Using coordinates in which $K^m = \delta ^m_1$ we deduce
that $y^1=\theta$ and the metric may be written as
\begin{eqnarray}
g_{mn} dy ^m dy ^n =d \theta ^2 + g _{m n}
 dy ^m  dy ^n   ,
\end{eqnarray}
where the summation on the right hand side is from 2 to
the dimension of $Y$ and  $g_{mn}$ is independent  of $\theta$.

We may apply these   results to either eleven or ten dimensional
supergravity. If we want $X$ to be Minkowski spacetime  and in addition we
want some supersymmetry, then $Y$ must be Ricci flat and
admit some covariantly constant spinor fields. If it is irreducible
and seven dimensional it must have holonomy  $G_2$
and if it is six dimensional it must have holonomy in $SU(3)$.
Note that one can in fact deduce that if $Y$ is irreducible
and that it  admits a
covariantly
constant spinor then it must be Ricci flat. We saw regardless of
whether we had Killing spinors, that $Y$ must be Ricci flat
and admit no continuous  isometries.

Note that in ten-dimensional string theory, as opposed to ten
dimensional supergravity, we have extra complications.
Yang-Mills fields are present which are not coupled minimally to
the metric but  in a more subtle way to the curvature
involving Chern-Simons terms. The consequence is that if we start
with a purely supergravity solution  and then identify  the Yang-Mills
connection with  (part of) the Levi-Civita  connection, then it  will
automatically satisfy the Yang-Mills equations of motion.
Moreover the source in  the dilaton equations of motion
vanishes , and it thus  is consistent to take the dilaton
to be constant, thus evading the No-Go theorem which would otherwise
follow. The full curvature modified
Einstein  equations are now much more complicated and
the simple arguments given above do not directly apply.
Thus in string theory
the Yang-Mills sector combined with the curvature contributions
does in effect  conspire to violate the Strong Energy
Condition. Similar remarks apply to M-theory, as opposed
to eleven-dimensional supergravity. If extra curvature terms
are introduced then the simple arguments given above need not apply.

Of course a No-Go Theorem is no better than the assumptions that go
into it. If one considers non-compact internal spaces
\cite{GibbonsHull} or couples super-matter to supergravity \cite{Fre}
then one can obtain de-Sitter space as a solution of the classical
equations
of motion.

\section{The Tachyon}

In the light of these rather discouraging results,
it seems clear that it is worth exploring
a new idea. Moreover it is very striking that an intrinsic
part of string theory, both open and closed, and  apparent from its  very
beginnings in Regge theory,  is the existence of  tachyons. It is true
that in closed string theory they are projected out, but even there
the projection mechanism has only been checked explicitly
in perturbation theory to 2 loops! In open string theory tachyons  abound.
Great progress has been made of late in understanding their
significance. I shall not attempt to describe any of that work
in detail. I shall merely take away from it the idea that
one the existence of a tachyon in the perturbative spectrum
indicates that the perturbative vacuum is unstable
and that there exist a true vacuum, with zero energy density,
toward  which a tachyon field $T(x)$ naturally moves. Moreover
it seems that aspects of this process can be capture by comparatively  simple
effective field theory models. Perhaps the simplest is that
proposed by Sen. It has (in units in which $2 \pi \alpha ^\prime =1$)
the Lagrangian
\begin{eqnarray}
L &=& -V(T)\sqrt{1+ g^{\mu \nu} \partial_\mu T \partial _\nu T}
  \sqrt{-\det ( g_{\mu \nu})  } \cr
  &=& -V(T) \sqrt{-\det \thinspace G_{\mu \nu}}, \cr \nonumber\
\end{eqnarray}

where what we shall call the {\it tachyon metric} is given by
\begin{eqnarray}
G_{\mu \nu} & =& \big(G \bigr ) _{\mu \nu} \cr & =&
 g_{\mu \nu} + \partial _\mu T\partial _\nu. T\nonumber\cr
\end{eqnarray}
The potential $V(T)$ is taken to be non-negative
have a unique local maximum at the origin
$T=0$ and a unique global minimum away from the origin at which $V$ vanishes.
In the most interesting case  the global minimum is taken to lie
at $|T| =\infty$. Obviously more complicated potentials may be
contemplated
but this is the simplest case to begin with. Small  fluctuations
around the ``false vacuum''  at $T=0$ have negative mass squared
and so  it is unstable. How does the system evolve?

The equation of motion is
\ben
\Bigl (g^ {\mu \nu} -{ \partial ^\mu T \partial ^\nu T \over 1 +
  (\partial T ) ^2 }  \Bigr ) \partial _\mu \partial _\nu T = - { V
  ^\prime \over V} \bigl ( 1+ (\partial T) ^2 \bigr ). \label{eom}
\een
From (\ref{eom}) we deduce that contrary to
popular prejudice:  {\sl the tachyon is not a tachyon}!
If we define the tachyon co-metric by
\ben
 \bigl ( G^{-1} \bigr ) ^{\mu \nu} = g^ {\mu \nu}
 -{ \partial ^\mu T \partial ^\nu T \over 1 +
  (\partial T ) ^2 },
\een so that
\ben
\bigl ( G^{-1} \bigl ) ^ {\mu \nu}  G_{\nu  \lambda } = \delta ^\mu _\lambda,
\een
we see that the {\it characteristic cones}  of (\ref{eom}) are given by
the co-metric $( G^{-1}) ^{\mu \nu} $
and the {\it rays} by the metric $G_{\mu \nu}$. It is these which
govern the speed of propagation of small disturbance around some
background,
as may  readily be seen by  linearising (\ref{eom}).
According to the general theory of hyperbolic partial differential
equations,
the characteristic cones also govern the maximum speed of any
possible signal. Note that the  characteristic
 cones do not depend in any way upon the potential
$V(T)$, but just on the coefficients of the highest  derivative term
in the equations of motion.  Thus for example linearizing around flat
spacetime, we see that the maximum signal speed is that of light
because $g_{\mu \nu }$  and $G_{\mu \nu}$  coincide in that case.
 However in general
they will not coincide with the standard Einstein cone $g_{\mu \nu}$
and co-cone $g^{\mu \nu}$, in other words, in a
non-trivial tachyon background,
tachyon fluctuations will travel at a different speed from that of
light. Nevertheless even in this case, because
\ben
G_{\mu \nu} l^\mu l^ \nu   = \bigr ( l ^\mu \partial _\mu T \bigl ) ^2,
\een
for any vector $l^\mu $ lying on the Einstein cone, i,e, such that
\ben
g_{\mu \nu} l ^\mu l ^\nu =0,
\een
it is clear that the Einstein cone  lies outside  or on  the tachyon
cone, which we  denote by
\ben
g_{\mu \nu }  \ge G_{\mu \nu} \label{greater}
\een
and no super-luminal propagation is ever possible.

The energy momentum tensor $T^{\mu \nu}$  of the tachyon takes the
 form:
\ben
T^{\mu \nu}
=-V \sqrt {1+ (\partial T) ^2 } \bigl (  G^{-1}  \bigr ) ^{\mu \nu}
\een
from  which one deduces from (\ref{greater}) that it satisfies
the dominant energy condition, with respect to the  Einstein
metric $g_{\mu \nu}$ (as well in fact  with respect to the
 metric $G_{\mu  \nu}$).
which according to a result of Hawking,
ensures that if the tachyon vanishes at time zero
 outside some compact set, then it vanishes outside the future of
that compact set.  Specifically, in a local  frame in which
$T$ depends only upon time,
\ben
\rho = { V(T ) \over \sqrt{   1- {\dot T} ^2}  }
\qquad P= -V(T)
\sqrt{1-{\dot T} ^2 } .
\een
Thus
\ben
P = -{V^2(T) \over \rho }. \label{Chaplygin}
\een
Note that because of the square root, $|{ \dot T}|$ can never exceed
unity. If it were the case that $V(T)$ were a constant, independent of
$T$
then (\ref{Chaplygin}) is the equation of state of what is called
a {\it  Chaplygin gas}. As a pedagogic warm up for the real thing, it is
interesting to review the cosmology of such a gas as worked out by
Karmenshik,
Moshella and Pasquier \cite{Chaplygin}. It is a special
case of what have been called ``k-essence'' cosmologies \cite{essence}.

\section{Chaplygin Cosmology}

We  set $P= -{A   \over \rho}, \qquad A>0$. The first law (\ref{First}) may be
integrated to give the density $\rho$ as a function of the scale
factor
$a$. One gets
\ben
\rho = \sqrt { A  + { B \over a^6} },
\een
where $B$ is an integration constant which we take to be positive.
Clearly, at large scale factor, i.e at late times, we have a
cosmological term \ben \rho =-P  = {\rm constant} .\een
At small scale  factor we have a  dust-like behaviour,
$P=0$, and
\ben\rho \propto { 1\over a^3}.\een
The scale factor can  thus make a smooth transition from
a matter dominated form $a\propto t^{2 \over 3}$ to
an exponentially  inflating form $a \propto \exp{  H_\infty t}$
with $H_\infty = \sqrt {{ 8 \pi G  \sqrt{A} \over 3} }$ .

At large scale factor we have corrections to the pure
cosmological term:
\ben
\rho \approx \sqrt{ A} + \sqrt { B \over 4 A } { 1\over a^6},
\een
\ben
P\approx  -\sqrt {A} + \sqrt{ B \over 4A} {1  \over a^6}.
\een
Thus at late times  there is a small admixture of stiff matter.

\section{Tachyons and the strong energy condition}

It is clear  that at high density, both the Chaplygin
and the tachyon case violate the strong energy condition.
In fact
\ben
\rho + 3P= { V \over \sqrt {1 - {\dot T} ^2 }}
 \bigl ( 3 {\dot T}^2 -2 \bigr ).
\een
Thus the strong energy condition fails if
$|{\dot T}| < \sqrt{2 \over 3}$.
\section{Coupling to gravity}

We assume that the relevant action is
\ben
\int d^4 x  \Bigl (
 { R \over 16\pi G }
 \sqrt{-\det g_{\mu \nu} } -
 V \sqrt {- \det G_{\mu \nu} } \Bigr ) \qquad {\rm plus \thinspace boundary
  \thinspace term}.
\een
Substituting in the Raychaudhuri and Friedman equations gives
\ben
{{\ddot a} \over a} = { 8 \pi G \over 3}  { V (T)  \over \sqrt{ 1
    -{\dot T} ^2 } } \bigl (1- { 3 \over 2} {\dot T} ^2  \bigr ),
\een
\ben
{ {\dot a } ^2 \over a^2 } + {k \over a^2 } = { 8 \pi G \over 3} { V
  \over \sqrt{1-{\dot T} ^2 } }.
\een
The equation of motion for the tachyon field is
\ben
{{\ddot T} \over {  1 -{\dot T} ^2  } } + 3 { {\dot a} \over a }  {\dot T} +
{V^{\prime} \over V} =0.
\een
The first law becomes
\ben
{\dot \rho} + { 3 {\dot a} \over a} {\dot T} ^2 \rho=0. \label{fall-off}
\een
Since $|{\dot T}| \le 1$ we deduce from (\ref{fall-off}).
that
\ben
{\dot{ (\rho a^3)} }\le 0. \label{dominate}
\een
Now let us suppose that we set $T$ off some where near the top of the
potential at positive $T$ and $0< {\dot T} < \sqrt{2 \over 3}$.
 Let also  suppose that $k=0$ and that
$V(T)$ takes its  minimum value of zero at infinity.
The tachyon  field will
increase monotonically, eventually reaching infinity with ${\dot
  T}=1$.
Initially the scale factor will accelerate but this acceleration is
self-limiting, eventually become negative  ( once $ {\dot T} > \sqrt{2
  \over 3}$) and slowing down like pressure free matter.
From  (\ref{dominate}) we deduce that the pressure falls dramatically
(at least  exponentially for reasonable choices of $V(T)$).

\section{Shortcomings}

Many authors have pointed out the shortcomings of
the rolling of the tachyon  as a
mechanism for inflation \cite{short}. They include
\begin{itemize}
\item The natural scale of the model is the Planck or String scale.
 Planck  scale inflation will in general give fluctuations which are
too large.
\item If the mass scale near the local maximum at the origin
 is too large,
 then insufficient inflation will result before the tachyon
field nears the minimum value of the potential.
\item The model requires fine tuning so as to avoid being becoming
matter dominated before there is time for a hot radiation
era  during which nucleosynthesis takes place. Alternatively,
if one wants the tachyon to act as
the cold dark matter apparently seen  at the present time,
again one seems to require an element of fine-tuning.

\item Because the tachyon does not oscillate in a potential well at
the end of rolling, the usual mechanism for reheating does not operate.
\end{itemize}

\noindent Although these are all perfectly valid objections, in my opinion,
 they ignore
\begin{itemize}
\item The provisional and approximate nature of the model.
\item The incomplete nature of the modelling of the relation between
the open and closed string sectors.
\end{itemize}

\noindent For these reasons, I feel that it is too early to abandon completely
the idea that the tachyon may have a role to play in cosmology,
particularly at very early times. However, as we shall see,
if those times were of Planck scale,  there
remain difficulties in accommodating any such model with the absence of
large
gravitational wave perturbations. Before passing to that, I wish to make
some pedagogic comments on the issue of fine tuning.

\section{Fine tuning, initial conditions and Anthropic Considerations}

Issues of fine tuning in cosmology, as opposed to particle physics,
are concerned with plausible initial conditions in real time rather
than the evolution of the renormalization group equations
with energy scale.
At the classical level, this requires
some sort of {\it a priori} probability distribution
on the space of Cauchy data. At the quantum level, some sort of {\it a
priori}
measure on the space of initial quantum  states, that is an {\it a
priori}
density matrix against which to test the plausibility
of various ``proposals''  for the wave function of the universe
\cite{Leo}.
To make this quantitative, rather than just a description of personal
prejudice, seems to be extremely difficult.

Even at the classical level, and for a finite number of degrees of
freedom,
such as we have in the model discussed above, this is an extremely
 tricky
business. We have in effect a ``mini-superspace`` model, with the
structure
of a Hamiltonian system with a constraint. The constraint  is  that the
flow
$\bigl (q(t),p(t) \bigr )$ in a $2n$-dimensional phase space $\cal P$
 is confined to the $(2n-1)$- dimensional  submanifold
$\Gamma$ on which the  Hamiltonian vanishes
\ben
H(p,q)|_\Gamma=0.
\een
Precisely because we have an autonomous  dynamical system on $\Gamma$
we know that given {\sl any} present condition, there must be {\sl some}
initial
conditions which gave rise to it, namely any point on the trajectory
or classical history through our present point.
Thus we cannot expect a complete
``chaotic'' style explanation of our present condition: that all initial
conditions will inevitably give rise to it. This would only be
mathematically possible if we consider what happens if the time went
strictly to infinity, but clearly this is not true in cosmology.
 Rather we need a
purely  classical
measure
on the possible histories. This has been provided to us by Liouville
and was introduced for this purpose in
\cite{ghs} (see  \cite{Henneaux0} for its independent introduction
and  use in the context of
Bianch-IX models).
We first seek to capture all the trajectories on $\Gamma$ by cutting
with a $2n-2$ dimensional surface $S $ intersecting each
trajectory
once and only once. $S$ is the space of  classical histories
and the point  is that the Liouville or symplectic measure
$d^np \thinspace d^nq$ on $\cal P$ descends to $S$ via
a construction called the
Symplectic or Marsden-Weinstein quotient.

Using this symplectic measure one may now break $S$ into
two domains, one, $S_{\rm yes}$,
 with histories  close enough to what we believe our own
to have been
and, those $S_{\rm no}$  unlike our own.

We can now evaluate the measure and hence probabilities ,
 according to Liouville. The problem
is
that typically both
\ben
{\rm measure} (S_{\rm yes}) =\infty
\een
and
\ben
{\rm measure} (S_{\rm no}) =\infty.
\een
In other words quantities like for example the
``probability
of sufficient inflation'' which are constructed form ratios of these
 infinite
quantities   are just not well defined \cite{HawkingPage,HollandsWald}.

This problem, which arises even in the simplest finite dimensional
case, and is essentially an infra-red problem having nothing to do
with the usual need to impose an ultra-violet cutoff in quantum
theory, gets much much worse if one goes to the full field theory
with its infinite number of degrees of freedom. Similarly, in the
quantum
theory, even of one particle, one arrives at the problem that one
cannot easily normalise density matrices, and
 certainly not the unit
density matrix, which represents complete ignorance.
It is for these reasons I am extremely skeptical of claims
that the {\it Anthropic Principle} can ever be elevated to the level
of a precise quantitative tool. On the other hand, the Anthropic
Principle
does seem to give a very convincing explantion of why the world
is $3+1$ -dimensional. Interesting renormalizable
quantum field theories of the sort
we use to describe intelligent life cease to exist in higher
dimensions.
This comment is particularly relevant to brane scenarios
in which our universe may contain branes of different dimensions.
All may have excitations propagating  on them described  by
quantum field theories
but only on 3-branes will Nobel prizes be awarded for
their discovery \cite{Meudon}.

\section{Tachyon Condensation}

The most optimistic scenario  is that
\begin{itemize}
\item All open string states get confined in the true
tachyon ground state.
\item Closed string states arise as flux tubes or some sort of related
solitonic or non-perturbative excitations.
\end{itemize}

To implement this idea is the ambition of many people.
It is  a major challenge in string theory.
 In what follows, I will offer a commentary on how this might look
like at the level of an effective classical field theory.
More details of the underlying microphysics may be found in
for instance \cite{Sen1,GHashimotoY}.

Since the precise tachyon Lagrangian  is  not completely
  known at present,
although there exist a number
of impressive  calculations, and in any case any actual  Lagrangian
may change under field redefinitions, we consider a general one
\ben
L=L(T, y),  \qquad y= g^{\mu \nu}\partial_\mu  T \partial_\nu  T.
\een
From a cosmological point of view such Lagrangians  correspond
to what is
called
``k-essence'', a pun on ``quintessence'' \cite{essence}.

\section{Carollian Confinement}

The confinement of open string states may be given a rather
elegant kinematic description in terms of an old, but hitherto
unused, idea.

The Poincar\'e group $E(n-1,1)$,\footnote{ sometimes referred to as the
inhomogeneous Lorentz group and  such written as $ISO(n-1,1)$}
has two important In\"on\"u-Wigner contractions which occur in
the limiting cases when $c
\uparrow \infty$ or $\downarrow 0$ . The former case corresponds to
the Galilei group when we have instantaneous propagation and
action at a distance with fields satisfying elliptic partial
differential equations, the latter, which is less well known, is
called the Carroll group \cite{Leblond, Sen} and corresponds to
the case of no propagation at all. Fields at each spatial point
evolve independently and are typically governed by ordinary
differential equations with respect to just the time variable. For
that reason, this case often arises as the symmetry group of an
approximation scheme in which spatial derivatives are ignored
compared with time derivatives. Such approximation schemes are
sometimes called ``velocity dominated''.

Geometrically the Galilei group arises when the future light cone
flattens out to become a spacelike hyperplane. The Carroll group
arises when it collapses down to a timelike half line. In the
Galilean case only the contravariant metric tensor has a well
defined limit as $c \uparrow \infty$:

\ben \eta ^{\mu \nu} \rightarrow {\rm diag} (0,1,1,\dots ,1)
 \label{Galilei}
 \een
and the limiting spacetime structure is called a Newton-Cartan
spacetime. In the case of the Carollian limit it is the covariant
metric tensor which survives

\ben \eta_{\mu \nu } \rightarrow {\rm diag} (0,1,1,\dots, 1),
 \label{Carroll} \een
and one has a Carrollian spacetime.

Now let's look at Sen's tachyon metric in the limit of tachyon
condensation,
i.e. in the limit that $|{\dot T} | \rightarrow 1$.
We have
\begin{eqnarray}
G_{\mu \nu} &=& g_{\mu \nu} + \partial _\mu T \partial _\nu T \cr
            &=& {\rm diag} \bigr( -1+ {\dot T} ^2 , 1,1,1  \bigl ) \cr
            &\rightarrow & {\rm diag}\bigr ( 0,1,1,1 \bigl )\cr\nonumber
\end{eqnarray}
Clearly as $|{\dot T}| \rightarrow 1$, the  cone defined by
$G_{\mu \nu}$ squeezes onto a half-line and no open tachyonic
excitations can propagate. If one thinks of these as sound waves,
then the speed of sound goes to zero and hence the pressure drops to
zero.

In fact, one may consider a more general tachyon Lagrangian, possibly coupled
to a Born-Infeld vector field $A_\mu$ with field strength $F_{\mu
\nu}=\partial _\mu A_\nu - \partial _\nu A _\mu$.
\ben
L= -V(T) \sqrt{-\det ( g_{\mu \nu} + F_{\mu \nu} ) }  \thinspace {\cal F} (z),
\een
with
\ben
z= \bigr ( G ^{-1}_{\rm open }\bigl ) ^{\mu \nu} \partial _\mu T \partial _\nu T,
\een
where the open string co-metric is given by
\ben
 \bigr ( G ^{-1}_{\rm open }\bigl ) ^{\mu \nu}= \Bigl( { 1 \over g+F }
 \Bigr )^{(\mu \nu)}.
\een
We  know that in general, the open string metric $\bigl (G_{\rm open}
\bigr )_{\mu \nu} $lies inside or on the
closed string metric $g_{\mu \nu}$, coinciding along two special
null directions $l^\mu$ such that $F_{\mu \nu} l^\nu =0$.
It also seems to satisfy a version of the Equivalence Principle,
it is  {\sl universal}  for all open string states, just as the closed
string metric satisfies the standard Einstein Equivalence Principle:
it is universal for all closed string states.

For the time being lets set $F_{\mu \nu}=0$.
We then have in Sen's case
\ben
L=V(T) \sqrt{1+y},
\een
or from boundary string conformal field theory (BSFT)
\ben
L= e^{-{ 1\over 2} T^2 } {\cal F}(y),
\een
with
\ben
{\cal F} (y) = { y 4^y \Gamma(y)^2 \over 2 \Gamma (2y) }.
\een
In general the energy momentum tensor is
\ben
T^{\mu \nu}= L g ^{\mu \nu} -2 L_y \partial ^\mu T \partial ^\nu T,
\een
thus
\ben
\rho =2y L_y -L,  \qquad P=L.
\een
The dominant energy condition will hold as long as
\ben
2yL_y -L \ge 0.
\een
The equations of motion are
\ben
\bigl( G^{-1}  \bigr )^{\mu \nu} \partial _\mu \partial _\nu T = {
L_T\over 2L_y },
\een
and the propagation  co-metric given by
\ben
\bigl( G^{-1}  \bigr )^{\mu \nu}=
g^{\mu \nu} + { 2 L_{yy} \over L_y} \partial^\mu T \partial ^\nu T,
\een
and the propagation metric given by
\ben
G_{\mu \nu}= g_{\mu \nu} - { 2 L_{yy} \over L_y + 2 L_{yy} }.
\een
Thus we shall get causal propagation if
\ben
{2 L_{yy} \over L_y} \le 0.
\een

Using these formulae one may check that
not only Sen's energy momentum tensor but also that coming from
BSFT satisfy both the strong energy condition and have  causal
propagation. More importantly for the present considerations,
one may check that as we approach the condensate, $y \rightarrow -1$,
the metric becomes Carollian. In general we have
\ben
G_{\mu \nu } ={\rm diag}  \Bigl ( -1 -{\dot T} ^2 \thinspace { 2 Lyy
    \over L+y + 2 L_yy},1,1,1 \Bigr ).
\een
It follows  that as long as $|L_{yy}| \rightarrow \infty $ as $y
\rightarrow -1$ then
\ben
G_{\mu \nu } ={\rm diag }\Bigl ( 0 ,1,1,1 \bigr ).
\een
In fact in  the case of BSFT,
\ben
L \rightarrow - { 1 \over 2} {1 \over ( 1+y)}.
\een

It is an attractive extrapolation
 from this example to speculate  that, although the tachyon metric may not
enjoy the same universality properties as the open string metric
\cite{Herdeiro,West,Gibbons2,Gibbons3}, the
Carollian confinement property described here is universal.

\section{Inclusion of Fluxes}

What happens if we consider  a case in which $F_{\mu \nu} \ne 0$?
One way is to generalise Sen's action to
\ben
L=V(T) \sqrt{-\det ( g_{\mu \nu}  + F _{\mu \nu} + \partial _\mu T
    \partial _\nu T )  }
\een
In this case the tachyon field is on the same footing as a transverse
scalar
in the Dirac-Born-Infeld action for a brane.
Alternatively  one could  use the full BSFT action given above. In both cases
we look for a solution
with a constant electric field and find that the condensed  state at
$V(T) \rightarrow 0$  is now
given by
\ben
{\dot T} ^2 + E ^2 =1,
\een
with $E=|{\bf E}|$.
To understand the dynamics it is, as in a previously studied case,
convenient to pass to the Hamiltonian formulation of the theory.
Define the conjugate variables
\ben
{\bf D}= {\partial L \over \partial {\bf E} }, \qquad P= {\partial L
  \over \partial
{\dot T} } .
\een
The Hamiltonian density is
\ben
{\cal H} = {\bf D } .{\bf E } + P {\dot T} -L.
\een
In the Sen case one discovers that
\ben
{\cal H}=\sqrt {{\bf D} ^2 + P^2 + ( {\bf D} .\nabla T )^2 + ( F_{ij} D_j +
    \partial _i T ) ^2 + V^2 \det ( \delta _{ij} + F_{ij} + \partial
    _i T \partial _j T ) } \een
Obviously there is a smooth $V\downarrow 0$ limit.

One may now investigate the propagation
of small fluctuations in the limit when ${\dot T} ^2 + E^2 =1$.
One finds that
\begin{itemize}\item
Propagation in directions orthogonal to $ {\bf E}$ is suppressed.
\item Propagation along the direction of parallel to ${\bf E}$
has speed $\pm E$. This is just what one expects of a
fluid of parallel flux tubes or strings and is
consistent with earlier work on string fluids
 \end{itemize}
We recover that system by dropping the tachyon.
One gets, in the limit $V\downarrow 0$
\ben
{\cal H} =\sqrt {{\bf D} ^2 + ({\bf D} \times {\bf B})^2}.
\een
One should note that the energy of a static electric
flux line with ${\bf B}=0$ is proportional to its length, as expected,
because in that case ${\cal H} = |{\bf D} |$.
Now,
\ben
{\bf H} = { {\bf B} {\bf D} ^2 -{\bf D} ({\bf B}
 .{\bf D}) \over \sqrt { {\bf D} ^2 + ({\bf D} \times {\bf B})^2}}
\een
and
\ben
{\bf E} = { { \bf D} + {\bf D} {\bf B} ^2 - {\bf B} ( {\bf B} .{\bf D}
  \over \sqrt {{\bf D} ^2 + ({\bf D} \times {\bf B})^2}
}
\een
It follows that
\ben
{\bf D} .{\bf H}=0, \qquad {\bf D} ^2 - {\bf H} ^2 >0.
\een
Thus if one constructs an Amp\`ere tensor $K= {1 \over 2}  K_{\mu \nu} dx ^\mu
\wedge dx ^\nu $ from
$({\bf D}, {\bf H})$ in the same way that one construct the Faraday
tensor $F_{\mu \nu} $ from ${(\bf E}, {\bf B} )$ one gets.
 \ben
\det K_{\mu \nu} =0, \Leftrightarrow  K \wedge  K =0 ,
\een
and
\ben K_{\mu \nu } K^{\mu \nu } <0.
\een
Thus the two-form $K$ is simple and defines  a distribution
of timelike 2-planes
in the tangent space at each point of spacetime.
The field equations
\ben
dK=0,
\een
imply that this  distribution is integrable, i.e. that spacetime is
foliated by timelike 2-surfaces tangent to $K$. Physically
one may identify these surfaces with the world sheets of
a fluid of electric flux tubes. Their energy momentum tensor is given
by
\ben
T^{\mu \nu} = - { K^\mu \thinspace _\lambda K ^{\nu \lambda} \over
\sqrt{-{1\over 2} K_{\alpha \beta} K^{\alpha \beta}} }
\een
For a static flux tube
$$
T_{\mu \nu} =
\pmatrix
{
& {\cal H} & 0 & 0 & 0 \cr
& 0 &-{\cal H} & 0 & 0 \cr
& 0 & 0 & 0 & 0 \cr
& 0 & 0 & 0 & 0 \cr
 }  .
$$

This has a tension equal to the energy density in the direction
of the electric field and zero pressure transverse to the electric
field,
again as expected.

Although this classical model exhibits a string fluid behaviour
with flux tube solutions, there remain some obvious   problems.
Neither the thickness nor the value of the flux is determined.
Hopefully,  this will emerge in some future quantum mechanical treatment.

\section{ Carroll versus Galilei }

It is clear that there is some sort of duality between the
Carroll and Galilei
groups.
The purpose of this penultimate
section is  to describe  this duality  in a geometrical
way by lifting there action  up to one higher spatial  dimension
and exhibiting both groups as subgroups of the Poincar\'e
group $E(n,1)$. By passing to one further temporal dimension
one may embed the Poincar\'e group in the Conformal group
$E(n,2)$. The description of the Carroll group
I am about to describe
is not new, in that it is very briefly described in
 \cite{Duval}, which
 is largely about the lifting of the action of the Galilei group
to one  higher dimension first given by Kunzle and Duval (see
\cite{Duval} for original references). The reason for
returning to the subject here is the hope that it may afford
some more insight into the properties of the tachyon condensate.

In fact, as we shall see from our discussion,
the Carroll group will also emerge naturally
in brane-dymanics in  the limit that the brane world volume becomes
lightlike.
Another  direct consequence of our
analysis, is that one sees  the Carroll
group  emerging naturally  in the isometry group of certain pp-wave spacetimes
which have been intensively studied of late.
Much earlier the possible role of Carrollian spacetimes
near spacetime singularities and in the so called strong coupling
(i.e. large inverse gravitational tension  $G\over c^4$ \cite{Max}) limit of
General Relativity has been discussed by Henneaux \cite{Henneaux1, Henneaux2}
and this is closely related to recent  work on Kac-Moody symmetries
in M-theory \cite{Nicolai}. Finally  one might hope that
just as the Galilei-covariant theories can easily  be constructed
using the null reduction described below, so one might hope to
construct Carroll-covariant theories using the dual related ideas.
This might lead to an extension of the work in \cite{gomis}
\footnote{I am  grateful to Joaquim
Gomis for drawing my attention to the possible
interest of Carrollian string theories}.

\subsection{The  Lift}

Let us start with the well studied Galilei case. The basic idea is
to start with  flat Minkowski spacetime ${\Bbb E} ^{n,1}$ whose metric
written in
double null coordinates  $(u,v,x^i)$, $i=1,2,\dots, n-1$,  is
\ben
ds^2 =- 2 dudv +  dx^i d x^i.
\een

The Lie algebra of the Poincar\'e group $\frak{e}(n,1)$ is spanned by
the Killing vector fields generating the Lie algebra of the Euclidean
group $\frak{e}(n-1)$, translations and rotations
\ben
P_i = \partial _i \qquad  L_{ij}= x_i \partial _j - x_j \partial _i,
\een
two null translations and one boost
\ben
U=\partial _u, \qquad V= \partial _v \qquad N= u \partial _u -v \partial _v,
\een
and two further sets of boosts

\ben
U_i= u\partial _i + x _i \partial _v \qquad V_i=  v \partial _i + x_i
\partial _u.
\een
There is an obvious symmetry under inter-changing $u$ and $v$ induced
by reflection in the timelike $(n-1)$- plane $u=v$.

To obtain the Bargmann group, the central extension of the
Galilei group,
we ask for the subgroup which
commutes with the null translation generated by $V=\partial _v$.
This is generated by $\{ P_i, L_{ij}, U, V, U_i,   \}$. The Galilei
group is obtained by taking the quotient by the null translation group
$\Bbb R $ generated by $V$.
It is easy to see
that the Galilei group acts on the quotient  ${\Bbb E}^{n,1} /{\Bbb
 R}$ which may be identified with a  Newton-Cartan spacetime $M^n $,
the coordinate $u$ playing the role of Newtonian absolute time.
The generators $V_i$ are Galilean boosts. Because
\ben
[P_i, U_j ]= \delta_{ij} V, \label{Heisenberg}
\een
they commute with
spatial translations (modulo $V$)
but not with time translations
\ben
[U, U_i]=P_i.
\een

We may think of this construction in terms of a Kaluza-Klein type reduction
in which we think of ${\Bbb E} ^{n,1}$ as a fibre bundle
with projection map \ben \pi: {\Bbb E} ^{n,1} \rightarrow M^n \label{submersion}\een
given by $ (u,v,x^i)  \rightarrow (u, x^i)$.
However in contrast to the usual case, the fibres are lightlike.
Using the map $\pi $ one may push forward the Minkowski  co-metric
on ${\Bbb E} ^{n-1}$ down to the Newton--Cartan
spacetime to give the  degenerate co-metric (\ref{Galilei}).

To obtain the Carroll group, we ask instead for the subgroup
of the Poincar\'e group which leaves invariant the null hyperplane
$u={\rm constant}$. This is generated by $\{P_i, L_{ij}, V, U_i \}$.
Now the null coordinate $v$ plays the role of time. The Carollian
boosts
$U_i$ commute with time translation
\ben
[V, U_i]=0,
\een
 but by (\ref{Heisenberg}) they  no longer commute
with spatial translations $P_i$. In fact one obtains
a  Heisenberg sub-algebra
with the time translations being central.
From an algebraic point of view the Carrol and Galilei groups
differ only in the choice of generator of time translations: one picks
either $V$ or $U$.

One may think of the null hyperplane $u={\rm constant}$ as
the image under the embedding map \ben x: M^n \rightarrow {\Bbb E}
^{n,1} \label{immersion}, \een such that $(v,x^i) \rightarrow ({\rm
  constant},v,x^i)$,
of a Carollian spacetime time. The pull back of the Minkowski metric
gives the degenerate Carrollian metric (\ref{Carroll}).
Thus the duality relating the cases  is between an  immersion
$x$ (\ref{immersion})
and a submersion $\pi$  (\ref{submersion}) and interchanges
domain and range.

\subsection{Plane  Waves}

One may generalize the idea of a Newton-Cartan spacetime
to include gravitational fields by replacing  ${\Bbb E} ^{n,1}$
in the construction above by an $n+1$ dimensional spacetime admitting
a covariantly constant null Killing field. In the context
of Galilei kinematics such spacetime is said to admit
a Bargmann structure. Physically it corresponds to a plane-fronted
gravitational wave with parallel rays, or pp-wave for short. In harmonic
coordinates,
the metric takes the form
\ben
ds^2 = -2 du d{\tilde v} + H({\tilde x} , u) du ^2 + d{\tilde x}^i
d {\tilde x}^i.
\een
In the special case that the function $H({\tilde x}, u)$ is quadratic
in $\tilde x$, the isometry group is enhanced  from ${\Bbb R}$
to a $(2n+1)$-dimensional Heisenberg group. This acts on null
hypersurfaces
and  is a subgroup of the Carroll group. Gravitational
waves of this type are called plane waves and as emphasised by Bondi,
Pirani and Robinson in the case $n=3$  the symmetry group and
number of polarisation states  coincide exactly with what one obtains
from linearised theory and also for plane electromagnetic waves.
If $n>3$ the symmetry groups also coincide but the number of
polarisation states
of course differ.

\section{Concluding Remarks}
Given the speculative
picture of the tachyon condensate outlined above,
it is clear that issues such as ``reheating ''
and gravitational wave production may be very different
from what they are in the standard picture.
It therefore seems to me to be premature to rule out  a
r\^ole for the tachyon in cosmology.

Consider for instance, a world which initially contains both open
string states  and hence necessarily  closed string states. One could
envision an initial  {\it Open String Era} during which the
system rolled down to the true vacuum, a tachyon condensate
in which all open string states suffer {\it Carrollian Confinement},
there propagation cone collapses onto a half-line and thereafter only
closed string states can propagate. We now enter the {\it Closed
 String Era} in which presumably the open strings eventually reassemble
themselves to give the standard model during some {\it Primordial
Radiation Era} . After this point a more or less conventional
inflationary scenario could have set in.

 One might think that
we might never have access to any information form before
the confining phase transition separating the Open String Era from the
Closed String Era.  However this may not be completely correct.
As I   described earlier, we can treat in a rough way the
cosmological rolling to wards the tachyon condensate using a simple
FLRW model. We can also consider gravitational wave perturbations
around that background configuration. If these gravitational waves
can penetrate beyond the confining phase transition, then they should
be observable  today and this raises difficulties with the
observational data. This is of course just the basic problem with all
primordial
inflationary models.   However in this case things might just  be
different, because the graviton is just a closed   string state and
closed strings are supposed to be topological excitations, flux tubes in
the true
open string vacuum.  If that is true then the gravitons we can see
today must, in some sense have been created then. More particularly,
it is not obvious that any gravitons created earlier could have passed
through the Carrollian barrier. If that is true, then we need not
worry
about the present limits.

It appears  that a complete theoretical treatment
of  a scenario like this is way  out of reach of present  day techniques
in String Theory. Some relevant  new ideas involving
the Wheeler-De-Witt equation  may be found in \cite{Sen2}.
 Thus it must remain at present a speculation.
Nevertheless it seems to me well worth bearing  in mind, if only as
a challenge of our powers of theoretical analysis.

\end{document}